\documentclass{article}
\usepackage[utf8x]{inputenc}
\usepackage[a4paper]{geometry}
\usepackage{url}
\usepackage{bm}
\usepackage{amsmath}
\usepackage{amssymb}
\usepackage{graphicx}
\usepackage{setspace}
\usepackage[authoryear]{natbib}
\usepackage[unicode=true,pdfusetitle,
 bookmarks=true,bookmarksnumbered=false,bookmarksopen=false,
 breaklinks=false,pdfborder={0 0 1},backref=section,colorlinks=false]
 {hyperref}

\makeatletter

\providecommand{\tabularnewline}{\\}

\usepackage{float}\usepackage{mathrsfs}\usepackage{color}\usepackage{setspace}\usepackage{moreverb}

\usepackage[font=small,labelfont=bf]{caption}
\usepackage{afterpage}

\makeatother

\begin{document}

\title{Fast, Exact Bootstrap Principal Component Analysis for $p>$ 1 million}

\date{\today}

\author{Aaron Fisher, Brian Caffo, Brian Schwartz \& Vadim Zipunnikov}

\maketitle

\section*{Abstract }

Many have suggested a bootstrap procedure for estimating the sampling variability of principal component analysis (PCA) results. However, when the number of measurements per subject ($p$) is much larger than the number of subjects ($n$), calculating and storing the leading principal components from each bootstrap sample can be computationally infeasible. To address this, we outline methods for fast, exact calculation of bootstrap principal components, eigenvalues, and scores. Our methods leverage the fact that all bootstrap samples occupy the same $n$-dimensional subspace as the original sample. As a result, all bootstrap principal components are limited to the same $n$-dimensional subspace and can be efficiently represented by their low dimensional coordinates in that subspace. Several uncertainty metrics can be computed solely based on the bootstrap distribution of these low dimensional coordinates, without calculating or storing the $p$-dimensional bootstrap components. Fast bootstrap PCA is applied to a dataset of sleep electroencephalogram recordings ($p=900$, $n=392$), and to a dataset of brain magnetic resonance images (MRIs) ($p\approx3$ million, $n=352$). For the MRI dataset, our method allows for standard errors for the first 3 principal components based on 1000 bootstrap samples to be calculated on a standard laptop in 47 minutes, as opposed to approximately 4 days with standard methods.

\smallskip{}

\noindent
{\bf Keywords:}
functional data analysis, image analysis, singular value decomposition, SVD, PCA.

\section{Introduction }

Principal component analysis (PCA) \citep{jolliffe2005principal} is a dimension reduction technique that is widely used fields such as genomics, survey analysis, and image analysis. Given a multidimensional dataset, PCA identifies the set of basis vectors such that the sample subjects' projections onto these basis vectors are maximally variable. These new basis vectors are called the sample principal components (PCs), and the subjects' coordinates with respect to these basis vectors are called the sample scores. The sample PCs can be thought of as estimates of the population PCs, or the eigenvectors of the population covariance matrix. Consistency conditions for PCs in the high dimension, low sample size (HDLSS) context are discussed by \citet{jung_marron2009pca_consistency_HDLSS}, and depend on the spacing of the eigenvalues of the population covariance matrix. \citet{shen2013consistency_SPCA_HDLSS} expand this discussion to consistency conditions for sparse PCA, when the first eigenvector of the population covariance matrix can be assumed sparse. Consistency of the $n$-length, right singular vectors of high dimensional sample data matrices has been demonstrated by \citet{leek2011asymptotic}.

A fundamental drawback of the PCA algorithm is that it is purely descriptive -- there is no clear method for estimating the sampling variability of the scores, the PCs, or proportion of variance that each PC explains. Analytically derived, asymptotic confidence intervals for PCs typically require the assumption of normally distributed data \citep{girshick1939sampling,tipping1999probabilistic}, or existence and computation of fourth order moments which results in $O(p^{4})$ complexity \citep{kollo1993asymptotics,kollo1997asymptoticsHotelling,ogasawara2002concise_stderrs_PCs}, where $p$ is the sample dimension. As an alternative to analytical, asymptotic confidence intervals, \citet{diaconis1983sciAmerican} proposed bootstrap based confidence intervals for PCA results. The bootstrap has also been discussed in the context of factor analysis \citep{chatterjee1984factor,thompson1988program,lambert1991CIs}, and in the context of determining the number of nontrivial components in a dataset \citep{lambert1990assessing,jackson1993stopping,peres2005many,hong2006bootstrap}. However, when applying the bootstrap to PCA in the high dimensional setting, the challenge of calculating and storing the PCs from each bootstrap sample can make the procedure computationally infeasible. 

To address this computational challenge, we outline methods for exact calculation of PCA in high dimensional bootstrap samples that are an order of magnitude faster than the current standard methods. These methods leverage the fact that all bootstrap samples occupy the same $n$-dimensional subspace, where $n$ is the original sample size. Importantly, this leads to bootstrap variability of the PCs being limited to rotational variability within this subspace. To improve computational efficiency, we shift operations to be computed on the low dimensional coordinates of this subspace before projecting back to the original $p$-dimensional space. 

There has been very little work applying bootstrap to PCA in the high dimensional context, largely due to computational bottlenecks. The methods we propose drastically reduce these bottlenecks, allowing for simulation studies of PCA in high dimensions, and for further study of bootstrap PCA in real world, high dimensional scientific applications.

Our methods can also be directly applied to determine the resampling-based variability of any model that depends on a singular value decomposition of the sample data matrix. Examples include bootstrap and cross-validation variability for principal component regression (PCR), independent component analysis (ICA), ridge regression, and, more generally, regression with quadratic penalties.

The remainder of this paper is organized as follows. Section \ref{sub:brief-summary-notation} presents some initial mathematical notation, and gives a basic summary of PCA and the bootstrap procedure. Section \ref{sub:Fast-BootPCA-intuition} outlines the intuition for fast bootstrap PCA. Section \ref{sec:Motivating-Data} discusses two motivating data examples -- one based on sleep electroencephalogram (EEG) recordings, and one based on brain magnetic resonance images (MRIs). Section \ref{sub:Additional-details-of} presents the full details of our methods for fast, exact bootstrap PCA. The computation complexity of our methods depends on the final sampling variability metric of interest. For example, pointwise standard errors for the PCs can be calculated more quickly than the full, high dimensional bootstrap distribution of the PCs. Section \ref{sec:Simulations-of-CI} uses simulations to demonstrate coverage rates for confidence regions around the PCs. Section \ref{sec:apply-FEB-PCA} applies fast bootstrap PCA to the EEG and MRI datasets, and presents computation times for different levels of sample size and sample dimension.

\subsection{A brief summary of PCA, SVD, the bootstrap, and their accompanying notation\label{sub:brief-summary-notation}}

In the remainder of this paper, we will use the notation $\mathbf{X}_{[i,k]}$ to denote the element in the $i^{th}$ row and $k^{th}$ column of the matrix $\mathbf{X}$. The notation $\mathbf{X}_{[,k]}$ denotes the $k^{th}$ column of $\mathbf{X}$; $\mathbf{X}_{[k,]}$ denotes the $k^{th}$ row of $\mathbf{X}$; $\mathbf{X}_{[,1:k]}$ denotes the first $k$ columns of $\mathbf{X}$; and $\mathbf{X}_{[1:k,1:k]}$ denotes the block of matrix $\mathbf{X}$ defined by the intersection of the first $k$ columns and rows. The notation $\mathbf{v}_{[j]}$ denotes the $j^{th}$ element of the vector $\mathbf{v}$, the notation $\mathbf{1}_{k}$ denotes the $k$-length vector of ones, and the notation $\mathbf{I}_{k}$ denotes the $k\times k$ identity matrix. We will also generally use the term ``orthonormal matrix'' to refer to rectangular matrices with orthonormal columns.

In order to create highly informative feature variables, PCA determines the set of orthonormal basis vectors such that the subjects' coordinates with respect to these new basis vectors are maximally variable \citep{jolliffe2005principal}. These new basis vectors are called the sample principal components (PCs), and the subjects coordinates with respect to these basis vectors are called the sample scores.

Both the sample PCs and sample scores can be calculated via the singular value decomposition (SVD) of the sample data matrix. Let $\mathbf{Y}$ be a full rank, $p\times n$ data matrix, containing $p$ measurements from $n$ subjects. Suppose that the rows of $\mathbf{Y}$ have been centered, so that each of the $p$ dimensions of $\mathbf{Y}$ has mean zero. The singular value decomposition of $\mathbf{Y}$ can be denoted as $\mathbf{VDU}'$, where $\mathbf{V}$ is the $p\times n$ matrix containing the orthonormal left singular vectors of $\mathbf{Y}$, $\mathbf{U}$ is the $n\times n$ matrix containing the right singular vectors of $\mathbf{Y}$, and $\mathbf{D}$ is a $n\times n$ diagonal matrix whose diagonal elements contain the ordered singular values of $\mathbf{Y}$. The principal component vectors are equal to the ordered columns of $\mathbf{V}$, and the sample scores are equal to the $n\times n$ matrix $\mathbf{DU}'$. The diagonal elements of $\left(1/(n-1)\right)\mathbf{D}^{2}$ contain the sample variances for each score variable, also known as the variances explained by each PC. Approximations of $\mathbf{Y}$ using only the first $K$ principal components can be constructed as $\hat{\mathbf{Y}}:=\sum_{k=1}^{K}\mathbf{V}_{[,k]}(\mathbf{DU}')_{[k,]}$. Existing methods for fast, exact, and scalable calculation of the SVD in high dimensional samples are discussed in the supplemental materials.

The sampling variability of PCA can be estimated using a bootstrap procedure. The first step of this procedure is to construct a bootstrap sample, by drawing $n$ observations, with replacement, from the original demeaned sample. PCA is reapplied to the bootstrap sample, and the results are stored. This process is repeated $B$ times, until $B$ sets of PCA results have been calculated from \textbf{$B$ }bootstrap samples. We index the bootstrap samples by the superscript notation $b$, so that $\mathbf{Y}^{b}$ denotes the $b^{th}$ bootstrap sample. Variability of the PCA results across bootstrap samples is then used to approximate the variability of PCA results across different samples from the population. Unfortunately, recalculating the SVD for all $B$ bootstrap samples has a computation complexity of order $O(Bpn^{2})$, which can make the procedure computationally infeasible when $p$ is very large.

\subsection{Fast bootstrap PCA -- resampling is a low dimensional transformation\label{sub:Fast-BootPCA-intuition}}

It's important to note that the interpretation of principal components (PCs) depends on the coordinate vectors on which the sample is measured. Given the sample coordinate vectors, the PC matrix represents linear transformation that aligns the coordinate vectors with the directions along which sample points are most variable. When the number of coordinate vectors $(p)$ exceeds the number of observations $(n)$, this transformation involves first reducing the coordinate vectors to a parsimonious, orthonormal basis of $n$ vectors whose span still includes the sample datapoints, and then applying the unitary transformation that aligns this basis with the directions of maximum sample variance. The first step, of finding a parsimonious basis, is more computationally demanding than the alignment step. However, if the number of coordinate vectors is equal to the number of datapoints, then the transformation obtained from PCA consists of only an alignment.

The key to improving computational efficiency of PCA in bootstrap samples is to realize that all resampled observations are contained in the same low dimensional subspace as the original sample. Because the span of the principal components $\mathbf{V}$ includes all observations in the original sample, the span of $\mathbf{V}$ also includes all observations in any bootstrap sample. Thus, in each bootstrap sample, $\mathbf{Y}^{b}$, we can skip the computationally demanding dimension reduction step of the PCA by first representing $\mathbf{Y}^{b}$ in terms of the parsimonious, orthonormal basis $\mathbf{V}$. Viewing the bootstrap procedure as a loop operation over several bootstrap samples, we see that the low dimensional subspace on which all sample points lie is loop invariant.

To translate this intuition into the calculation of the SVD for bootstrap samples, we first note that $\mathbf{Y}^{b}$ can be represented as $\mathbf{Y}\mathbf{P}^{b}$, where $\mathbf{P}_{[i,j]}^{b}=1$ if $\mathbf{Y}_{[,j]}^{b}=\mathbf{Y}_{[,i]}$ and zero otherwise. In each bootstrap sample, we then calculate its SVD, denoted by $\mathbf{V}^{b}\mathbf{D}^{b}\mathbf{U}^{b'}$, via the following steps\\

\begin{tabular}{rll}
$\mathbf{Y}^{b}$ & $=\mathbf{Y}\mathbf{P}^{b}$ & where $\mathbf{P}^{b}$ represents a resampling operation\tabularnewline
 & $=\mathbf{V}\mathbf{DU}'\mathbf{P}^{b}$ & where $\mathbf{DU}'\mathbf{P}^{b}$ is the matrix of resampled scores\tabularnewline
 & $=\mathbf{V}(\mathbf{A}^{b}\mathbf{S}^{b}\mathbf{R}^{b'})$ & where $\mathbf{A}^{b}\mathbf{S}^{b}\mathbf{R}^{b'}:=svd(\mathbf{DU}'\mathbf{P}^{b})$\tabularnewline
 & $=(\mathbf{V}\mathbf{A}^{b})\mathbf{S}^{b}(\mathbf{R}^{b})'$ & where ($\mathbf{V}\mathbf{A}^{b}$) and ($\mathbf{R}^{b}$) are orthonormal, and $\mathbf{S}^{b}$ is diagonal\tabularnewline
 & $=svd(\mathbf{Y}^{b})$ & \tabularnewline
\end{tabular}

Rather than directly decomposing the $p$-dimensional bootstrap sample $\mathbf{Y}^{b}$, we reduce the problem to a decomposition of the $n$-dimensional resampled scores, $svd(\mathbf{DU}\mathbf{P}^{b})=:\mathbf{A}^{b}\mathbf{S}^{b}\mathbf{R}^{b'}$. Because $\mathbf{V}$ and $\mathbf{A}^{b}$ are both orthonormal, their product $\mathbf{V}\mathbf{A}^{b}$ is orthonormal as well. Since $\mathbf{S}$ is diagonal and $\mathbf{R}^{b}$ is orthonormal, $(\mathbf{V}\mathbf{A}^{b})\mathbf{S}^{b}(\mathbf{R}^{b})$ is equal to the SVD of $\mathbf{Y}^{b}$. The singular values, and right and left singular vectors of the $\mathbf{Y}^{b}$ can then be written respectively as $\mathbf{D}^{b}=\mathbf{S}^{b}$, $\mathbf{U}^{b}=\mathbf{R}^{b}$, and $\mathbf{V}^{b}=\mathbf{V}\mathbf{A}^{b}$. If only the first $K$ principal components are of interest, then it is sufficient to calculate and store $\mathbf{A}^{b}$, \textbf{$\mathbf{U}^{b}$}, and $\mathbf{D}^{b}$ as the matrices containing only the first $K$ singular vectors and values of $\mathbf{DU}'\mathbf{P}^{b}$. Full details of our proposed methods for bootstrap PCA are discussed in section \ref{sub:Additional-details-of}.

\citet{daudin1988stability} applied an equivalent result to eigen-decompositions of bootstrap covariance matrices in the $p<n$ setting, but this result has not been widely used, nor has it been generalized to the $p>>n$ setting. \citet{daudin1988stability} suggested that, rather than decomposing the $p\times p$ covariance matrix, a more computationally efficient approximation is to decompose the covariance matrix of the $k$ leading resampled score variables. The eigenvectors of this $k\times k$ covariance matrix can then be projected onto the $p$-dimensional space to approximate the eigenvectors of the full $p\times p$ covariance matrix. In the $p>>n$ setting, however, if $k$ is set equal to $n$, then the approximation becomes exact. Note also that in the $p>>n$ setting, it is the projection onto the $p$-dimensional space that is most computationally demanding step (computational complexity $O(KBpn)$), rather than the $n$-dimensional decompositions (computational complexity $O(KBn^{2})$).

To gain intuition for why that the columns of $\mathbf{V}\mathbf{A}^{b}$ are the principal components of $\mathbf{Y}^{b}$, note that the resampled scores, $\mathbf{DU}'\mathbf{P}^{b}$, are equivalent to the resampled data, $\mathbf{Y}^{b}$, expressed in terms in terms of the coordinate vectors $\mathbf{V}$. This implies that the principal components of the resampled scores, $\mathbf{A}^{b}$, give the transformation required to align the coordinate vectors of the scores, $\mathbf{V}$, with the directions along which the resampled scores are most variable. Applying this transformation yields $\mathbf{V}\mathbf{A}^{b}$ -- the bootstrap principal components in terms of the sample's original, native coordinate vectors.

Random orthogonal rotations comprise the only possible way that the fitted PCs can vary across bootstrap samples. Because of this, the bootstrap procedure will not be able to directly estimate PC sampling variability in directions orthogonal to the observed sample, not unlike how a bootstrap mean estimate must be a weighted combination of the observed datapoints. However, when the inherent dimension of the population is small, the sampling variability of the PCs will generally be dominated by variability in a handful of directions, and these directions will generally be well represented by the span of the bootstrap PCs. Variability in directions not captured by the bootstrap procedure will tend to be of a much smaller magnitude. 

The rotational variability of the bootstrap PCs is directly represented by the $\mathbf{A}^{b}$ matrices. More specifically, information about random rotations within the $K$ leading PCs is captured by the $\mathbf{A}_{[1:K,1:K]}^{b}$ block matrices, which show how much each of the $K$ leading bootstrap PCs weight on each of original $K$ leading components. When the majority of bootstrap PC variability is due to rotations within the $K$ leading PCs, the $\mathbf{A}_{[1:K,1:K]}^{b}$ matrices provide a parsimonious description of this dominant form of variability.

Decomposing $\mathbf{V}^{b}$ into an alignment operation, $\mathbf{A}^{b}$, applied to the original sample components, $\mathbf{V}$, can drastically reduce the storage and memory requirements required for the bootstrap procedure, making it much more amenable to parallelization. Using this method, we're able to store all the information about the variability of $\mathbf{V}^{b}$ only by storing the $\mathbf{A}^{b}$ matrices, which can later be projected onto the high dimensional space. Calculating the $\mathbf{A}^{b}$ matrices only requires the low dimensional matrices $\mathbf{DU}'$ and $\mathbf{P}^{b}$, and does not require either operations on the $p\times n$ matrix $\mathbf{Y}^{b}$, or access to the potentially large data files storing $\mathbf{Y}$. In the context of parallelizing the bootstrap procedure, this allows for minimal memory, storage, and data access requirements for each computing node. 

Furthermore, in many cases, it is not even necessary to calculate and store the $p$-dimensional components, $\mathbf{V}_{[,1:K]}^{b}$. Instead we can calculate summary statistics for the bootstrap distribution of the low dimensional matrices $\mathbf{A}^{b}$, and translate only the summary statistics to the high dimensional space. For example, we can quickly calculate bootstrap standard errors for $\mathbf{V}_{[,1:K]}$ by first calculating the bootstrap moments of $\mathbf{A}^{b}$, and projecting these moments back onto the $p$-dimensional space (see section \ref{sec:Bootstrap-Moments-of}). Joint confidence regions for the PCs can also be constructed solely based on the bootstrap distribution of $\mathbf{A}^{b}$ (see section \ref{sec:Construction-of-CIs}). Similar complexity reductions are available when calculating bootstrap distribution of linear functions of the components, such as the the arithmetic mean of the $k^{th}$ PC (i.e. $(1/p)\mathbf{1}_{p}'\mathbf{V}_{[,k]}^{b}$). For any bootstrap statistic of the form $\mathbf{q}'\mathbf{V}_{[,k]}^{b}=(\mathbf{q}'\mathbf{V})\mathbf{A}_{[,k]}^{b}$, where $\mathbf{q}$ is a $ $$p$-length vector, the $n$-length vector $\mathbf{q'V}$ can be pre-calculated, and the complexity of the bootstrap procedure will be limited only be $n$.

\section{Motivating data \label{sec:Motivating-Data}}

In this section we apply standard PCA to a dataset of sleep EEG recordings ($p$=900), and to a dataset of preprocessed brain MRIs ($p$=2,979,666). A bootstrap procedure is later applied in section \ref{sec:apply-FEB-PCA}, to estimate sampling variability for the fitted PCs.

There has been demonstrated interest in the population PCs corresponding to both datasets \citep{di2009multilevel,crainiceanu2010population,zipunnikov2011ravenspca,zipunnikov2011multilevel_RAVENS}. For our purposes, the functional EEG data form an especially useful didactic example, as the sample PCs are also functional, and easily visualizable. We include the MRI dataset primarily to demonstrate computational feasibility of the bootstrap procedure when dimension ($p$) is large.

\subsection{Sleep EEG \label{sub:-EEG-description}}

The Sleep Heart Health Study (SHHS) is a multi-center prospective cohort study, designed to analyze the relationships between sleep-disorder breathing, sleep metrics, and cardiovascular disease \citep{quan1997shhs}. Along with many other health and sleep measurements, EEG recordings were taken for each patient, for an entire night's sleep. An EEG uses electrodes placed on the scalp to monitor neural activation in the brain, and is commonly used to describe the stages of sleep. Our goal in this application is to estimate the primary patterns in EEG signal that differentiate among healthy subjects, and to quantify uncertainty in these estimated patterns due to sampling variability.

To reflect this goal, we selected a subsample of 392 healthy, comparable controls from the SHHS ($n=392$). Our sample contained only female participants between ages 40 and 60, with no sleep disordered breathing, no history of smoking, and high quality EEG recordings for at least 7.5 hours of sleep. In order to more easily register EEG recordings across subjects, only the first 7.5 hours of EEG data from each subject were used. Although the EEG recordings consist of measurements from two electrodes, we focus for simplicity only on measurements from one of these electrodes (from the left side of the top of the scalp).

To process the raw EEG data, each subject's measurements were divided into thirty second windows, and the proportion of the signal in each window attributable to low frequency wavelengths (0.8-4.0 Hz) was recorded. This proportion is known as normalized $\delta$ power ($NP_{\delta})$, and is particularly relevant to deep stage sleep (NREM Stage 3). The preprocessing procedure used here to transform the raw EEG data into $NP_{\delta}$ is the same as the procedure used by \citet{crainiceanu2009eeg}. A lowess smoother was then applied to each subject's $NP_{\delta}$ function, as a simple means of incorporating the assumption that the underlying $NP_{\delta}$ process is a smooth function. This preprocessing procedure resulted in 7.5 hours $\times$ (60 minutes / hour) $\times$ (2 thirty second windows / minute) = 900 measurements of $NP_{\delta}$ per subject ($p=900$). 

\begin{figure}[!t]
\includegraphics[scale=0.5]{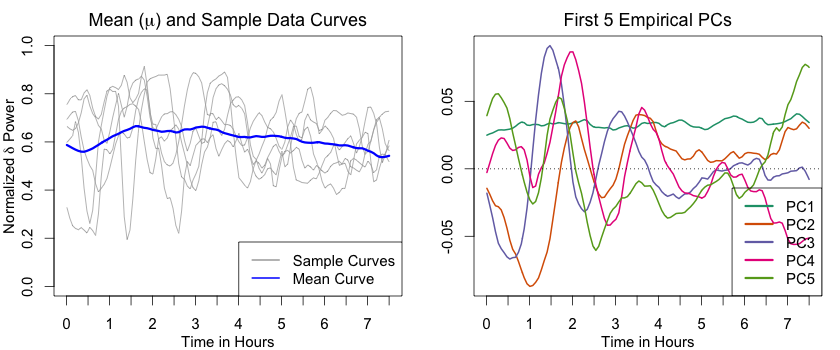}\caption{Summary of EEG dataset - The left panel shows examples of normalized $\delta$ power ($NP_{\delta}$) over the course of the night for five subjects, as well as the mean $NP_{\delta}$ function across all subjects (\textbf{$\bm{\mu}$}). The right panel shows the first five PCs of the dataset.\label{fig:Summary-of-EEG-data}}
\end{figure}

The left panel of Figure \ref{fig:Summary-of-EEG-data} shows examples of $NP_{\delta}$ functions for five subjects, as well as the mean $NP_{\delta}$ function across all subjects, denoted by \textbf{$\bm{\mu}$}. The first five principal components of the $NP_{\delta}$ data are shown in the right panel of Figure \ref{fig:Summary-of-EEG-data}. The first PC appears to be a mean shift, indicating that the primary way in which subjects differ is in their overall $NP_{\delta}$ over the course of the night. The remaining four PCs roughly correspond to different types of oscillatory patterns in the early hours of sleep. These components are fairly similar to the results found by \citep{di2009multilevel}, who analyze a different subset of the data, and employ a smooth multilevel functional PCA approach to estimate eigenfunctions that differentiate subjects from one another.

Collectively, the first five PCs explain approximately 55\% of the variation, and the first ten PCs explain approximately 76\% of the variation (see scree plot in supplemental materials). These estimates for the variance explained by each component are much lower than the estimates from \citet{di2009multilevel}. The difference is most likely due differences in how the MFPCA method employed by \citet{di2009multilevel} incorporates the assumption of underlying smoothness in $NP_{\delta}$.

\subsection{Brain magnetic resonance images\label{sub:Brain-Magnetic-Resonance}}

We also consider a sample data processed using voxel based morphometry (VBM) \citep{ashburner2000voxel_based_morphometry}, a technique that is frequently used to study differences in the size of brain regions across subjects, or within a single subject over time. Our data came from an epidemiological study of former organolead manufacturing workers \citep{stewart2006past,schwartz2007relations,schwartz2010evaluation,bobb2014cross}. We focused on the baseline MRIs from the 352 subjects for which both baseline and followup MRIs were recorded.

VBM images were constructed based on brain MRIs. The original MRIs were stored as 3-dimensional arrays, with each array element corresponding to tissue intensity in a voxel, or volumetric pixel, of the brain. Creating VBM images typically begins by registering each subject's brain MRI to a common template image, using a non-linear warping. The number of voxels mapped to each voxel of the template image during the registration process is recorded. This information is used to create subject-specific images on the template space, where each voxel's intensity represents the size of that voxel in the subject's original MRI. The VBM images used here were processed using a generalization of the regional analysis of volumes examined in normalized space (RAVENS) algorithm \citep{goldszal1998image,davatzikos2001vbmRAVENS}, and are the same as the baseline visit images used in \citep{zipunnikov2011multilevel_RAVENS,zipunnikov2011ravenspca}.

To create a single $p\times n$ data matrix, each subject's VBM image was vectorized, omitting the background voxels that did not correspond to brain tissue. The vector for each subject contained 2,979,666 measurements ($p$=2,979,666). Because the resulting data matrix was 3.5 Gb, it is difficult to store the entire data matrix in working memory, and block matrix algebra is required to calculate the sample PCs (see supplemental materials).

\begin{figure}
\includegraphics[scale=0.37]{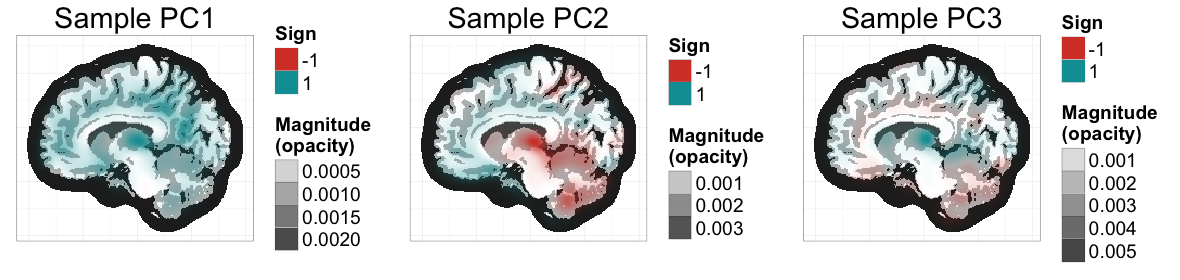}

\caption{MRI Sample PCs\label{fig:MRI-Sample-PCs}}

\end{figure}

A central slice from each of the first three PCs is shown in Figure \ref{fig:MRI-Sample-PCs}. The first PC appears to roughly correspond with grey matter, indicating that the primary way in which subjects regions tend to differ is in their overall grey matter volume. Together, the first 30 PCs explain approximately 53.3\% of the total sample variation (see scree plot in supplemental materials).

In the remainder of this paper, we refer to this dataset primarily as to demonstrate the computational feasibility of bootstrap PCA in especially high dimensions. Additional interpretation of the sample PCs is given in \citep{zipunnikov2011multilevel_RAVENS,zipunnikov2011ravenspca}.

\section{Full description of the bootstrap PCA algorithm\label{sub:Additional-details-of}}

In this section we outline calculation methods for bootstrap standard errors, bootstrap confidence regions, and for the full bootstrap distribution of the principal components (PCs). The overall computational complexity of the procedure depends on the bootstrap metric of interest, but the initial steps of all our proposed methods are the same.

Building on the notation of sections \ref{sub:brief-summary-notation} and \ref{sub:Fast-BootPCA-intuition}, let $K$ be the number of principal components that are of interest, which typically will be less than $n-1$. For simplicity of presentation, we assume that each dimension of the bootstrap sample $\mathbf{Y}^{b}$ has mean zero. Manually recentering $\mathbf{Y}^{b}$ however will not add any high dimensional complexity to the procedure, as this is equivalent to recentering the $n\times n$ matrix of resampled scores $\mathbf{DU}'\mathbf{P}^{b}$ (see supplemental materials).

For each bootstrap sample, we begin by calculating the leading $K$ singular vectors and singular values of the resampled scores $\mathbf{DU}'\mathbf{P}^{b}$. As noted in section \ref{sub:Fast-BootPCA-intuition}, the leading left and right singular vectors of $\mathbf{DU}'\mathbf{P}^{b}$ are stored as solutions for the $n\times K$ matrices $\mathbf{A}^{b}$ and $\mathbf{U}^{b}$ respectively. The leading singular values of $\mathbf{DU}'\mathbf{P}^{b}$ are the solutions for the diagonals of the $K\times K$ matrix $\mathbf{D}^{b}$. In the typical case where $K$ is less than the rank of $\mathbf{DU}'\mathbf{P}^{b}$, the first $K$ singular values of $\mathbf{DU}'\mathbf{P}^{b}$ are positive and unique, and the solutions for the columns of $\mathbf{A}^{b}$ and $\mathbf{U}^{b}$ are unique up to sign changes. Arbitrary sign changes in the columns of $\mathbf{A}^{b}$ will ultimately result in arbitrary sign changes in the bootstrap PCs. Adjusting for these arbitrary changes is discussed in section \ref{sub:Switching-the-sign}. 

We find in approximately 4\% of bootstrap samples from the MRI dataset, that although a solution to the SVD of $\mathbf{DU}'\mathbf{P}^{b}$ exists, the SVD function fails to converge. We handle these cases by randomly preconditioning the matrix $\mathbf{DU}'\mathbf{P}^{b}$, reapplying the SVD function, and appropriately adjusting the results to find the SVD of the original matrix. The full details of this procedure are described in the supplemental materials.

These baseline steps require a computational complexity of order $O(KBn^{2})$. They are sufficient for calculating the leading $K$ bootstrap scores and the variance explained by the leading $K$ bootstrap PCs.%
\footnote{The bootstrap score matrix is equal to $\mathbf{D}^{b}\mathbf{U}^{b'}$, and the variances explained by each bootstrap PC are equal to the diagonals of $(1/(n-1))(\mathbf{D}^{b})^{2}$. These variances explained can also be expressed as a proportions of the total variance of the bootstrap sample, which can be calculated as $trace(Var(\mathbf{Y}^{b}))=(1/(n-1))||\mathbf{DU}'\mathbf{P}^{b}||^{2}=(1/(n-1))\sum_{i=1}^{n}\sum_{j=1}^{n}(\mathbf{DU}'\mathbf{P}^{b})_{[i,j]}^{2}$. %
}

When moving on to describe the bootstrap distribution of the PCs, we have several options, each requiring a different level of computational complexity:
\begin{itemize}
\item \textbf{Standard errors} \textbf{for the PCs} can be calculated based on the bootstrap mean and variance of the columns of $\mathbf{A}^{b}$ (see section \ref{sec:Bootstrap-Moments-of}). These standard errors can be used to create pointwise confidence intervals (see section \ref{sub:Pointwise-Confidence-Intervals}). This option requires additional computational complexity of order $O(Kpn^{2}+KBn^{2})$.
\item \textbf{Joint confidence regions for the PCs} and for the principal subspace can be constructed using the methods in section \ref{sub:CRs-for-PS}. This option requires no additional computational complexity on the high dimensional scale.
\item \textbf{The full bootstrap distribution of PCs }can be calculated by projecting the principal components of the bootstrap scores onto the $p$-dimensional space (i.e. $\mathbf{V}_{[,1:K]}^{b}=\mathbf{V}\mathbf{A}^{b}$). The bootstrap PC vectors ($\mathbf{V}_{[,1:K]}^{b}$) can then be used to create pointwise percentile intervals for the PCs (see section \ref{sub:Pointwise-Confidence-Intervals}). If $p$ is sufficiently large such that the matrix $\mathbf{V}$ cannot be held in working memory, block matrix algebra can be used to break down the calculation of $\mathbf{V}\mathbf{A}^{b}$ into a series low memory operations (see supplemental materials). Calculation of all bootstrap PCs requires additional computational complexity of order $O(KBpn)$. If $K$ is set equal to $n-1$, then the computational complexity of this method is roughly equivalent to that the standard methods ($O(Bpn^{2})$). The total computation time, however, will still be approximately half the time of standard methods, as the matrices $\mathbf{Y}^{b'}\mathbf{Y}^{b}$ need not be calculated (see supplemental materials). 
\end{itemize}

\subsection{Adjusting for axis reflections of the principal components \label{sub:Switching-the-sign}}

Because the singular vectors of $\mathbf{Y}^{b}$ are not unique up to sign, arbitrary sign changes, also known as reflections across the origin, will induce variability in both the sampling and bootstrap distributions of the principal components ($\mathbf{V}^{b}$). These reflections, however, do not affect the interpretation of the PCs, and so their induced variability will cause us to overestimate sampling variability of the patterns decomposed by PCA \citep{mehlman1995comment,jackson1995reply,milan1995application}. For example, arbitrary sign changes can cause the confidence interval for any element of any principal component to include zero, even if the absolute value of that element is nearly constant and nonzero across all bootstrap samples.

To isolate only the variation that affects the interpretation of the PCs, we adjust the sign of the columns of $\mathbf{V}^{b}$ so that the dot products $\mathbf{V}_{[,k]}^{b'}\mathbf{V}_{[,k]}$ are positive for $k=1,2,...K$. Note that because $\mathbf{V}^{b}=\mathbf{V}\mathbf{A}^{b}$, sign changes in the columns of $\mathbf{V}^{b}$ are equivalent to sign changes in the columns of $\mathbf{A}^{b}$. For the same reason, sign adjustments for the columns of $\mathbf{V}^{b}$ are equivalent to sign adjustments for the columns of $\mathbf{A}^{b}$, which can be simpler to compute. Here, the dot products $\mathbf{V}_{[,k]}^{b'}\mathbf{V}_{[,k]}$ for $k=1,2....K$ actually do not require any additional calculations, as they can be found on the diagonal elements of $\mathbf{V}'\mathbf{V}^{b}=\mathbf{V}'\mathbf{V}\mathbf{A}^{b}=\mathbf{A}^{b}$. Independent of our work, this calculation simplification is also noted by \citet{daudin1988stability}. Whenever $\mathbf{A}_{[k,k]}^{b}$ is negative, we declare that an arbitrary sign change has occurred, and adjust by multiplying $\mathbf{A}_{[,k]}^{b}$ and $\mathbf{U}_{[,k]}^{b}$ by -1. The resulting PCs and scores are still valid solutions to the PCA algorithm.

Since $\mathbf{V}_{[,k]}^{b}$ and $\mathbf{V}_{[,k]}$ each have norm equal to one, their dot product is equal to the cosine of the angle between them. As a result, using the dot product $\mathbf{V}_{[,k]}^{b'}\mathbf{V}_{[,k]}$ to adjust for sign will ensure that the angle between $\mathbf{V}_{[,k]}^{b}$ and $\mathbf{V}_{[,k]}$ is between $-\pi/2$ and $\pi/2$. This range of angles is exactly the range that affects our interpretation of the bootstrapped PCs. Using these dot products for sign adjustment is also equivalent to choosing the sign of $\mathbf{V}_{[,k]}^{b}$ that minimizes the Frobenius distance $||\mathbf{V}_{[,k]}^{b}-\mathbf{V}_{[,k]}||$, a method that has been previously suggested \citep{lambert1991CIs,milan1995application}. 

It has also been suggested that the sign of each PC should be switched based on the correlation between the columns of $\mathbf{V}^{b}$ and the columns of $\mathbf{V}$, rather than the dot products $\mathbf{V}_{[,k]}^{b'}\mathbf{V}_{[,k]}$ \citep{jackson1995reply,babamoradi2012bootstrap}. While the two approaches are similar, and rarely give different results in practice, it is possible for the two methods to give different results when the PCs roughly correspond to mean shifts. We give examples of such cases in the supplemental materials, and argue that the results of the dot product method are more intuitive. In general, we find the dot product method to have a cleaner geometric interpretation than the correlation method.

\subsection{Bootstrap moments of the principal components\label{sec:Bootstrap-Moments-of}}

Traditional calculation of the mean and variance of $\mathbf{V}_{[,k]}^{b}$ requires first calculating the bootstrap distribution of $\mathbf{V}_{[,k]}^{b}$, and then taking means and variances over all $B$ bootstrap samples. However, using our characterization of $\mathbf{V}^{b}$ as $\mathbf{V}\mathbf{A}^{b}$, and properties of expectations, the same result can achieved without calculating or storing $\mathbf{V}_{[,k]}^{b}$. 

Specifically, the bootstrap mean $E(\mathbf{V}_{[,k]}^{b})$ can be found via $E(\mathbf{V}\mathbf{A}_{[,k]}^{b})=\mathbf{V}E(\mathbf{A}_{[,k]}^{b})$, where the operation $E$ is the expectation with respect to the bootstrap distribution. The bootstrap variance of $\mathbf{V}_{[i,k]}^{b}$ can be found via
\[
Var(\mathbf{V}_{[i,k]}^{b})=Cov(\mathbf{V}_{[,k]}^{b})_{[i,i]}=Cov(\mathbf{V}\mathbf{A}_{[,k]}^{b})_{[i,i]}=[\mathbf{V}Cov(\mathbf{A}_{[,k]}^{b})\mathbf{V}']_{[i,i]}=(\mathbf{V}_{[i,]})'Cov(\mathbf{A}_{[,k]}^{b})(\mathbf{V}_{[i,]})
\]

Where $Var$ and $Cov$ are variance operators with respect to the bootstrap distribution. The total computational complexity of finding $Cov(\mathbf{A}_{[,k]}^{b})$ and then $Var(\mathbf{V}_{[i,k]}^{b})$ for each combination of $i=1,2,...p$ and $k=1,...K$ is only $O(Kpn^{2}+KBn^{2}))$.%
\footnote{In practice, we calculate the diagonals of $\mathbf{V}Cov(\mathbf{A}_{[,k]}^{b})\mathbf{V}'$ by the row sums of $(\mathbf{V}Cov(\mathbf{A}_{[,k]}^{b}))\circ(\mathbf{V})$, where $\circ$ denotes element-wise multiplication as opposed to traditional matrix multiplication.%
}

This improvement in computation speed comes from pre-collapsing the complexity induced by having a large number of bootstrap samples before transforming to the high dimensional space. This allows us to separate calculations of order \textbf{$B$} from calculations of order $p$. Similar speed improvements are attainable whenever summary statistics or parametric models for the bootstrap distribution of $\mathbf{A}^{b}$ can be translated into summary statistics or parametric models for the high dimensional components $\mathbf{V}^{b}$.

\subsection{Construction of confidence regions\label{sec:Construction-of-CIs}}

Several types of confidence regions can be constructed based on the bootstrap distribution the PCs. In this section, we specifically discuss (1) pointwise confidence intervals (CIs) for the PCs, based on either the bootstrap moments or bootstrap percentiles; (2) confidence regions (CRs) for the individual PCs; and (3) CRs for the principal subspace. Only the pointwise percentile intervals require calculation of the full bootstrap distribution of the high dimensional PCs. All other CRs can be calculated solely based on the bootstrap distribution of the low dimensional $\mathbf{A}^{b}$ matrices.

\subsubsection{Pointwise confidence intervals for the principal components\label{sub:Pointwise-Confidence-Intervals}}

The simplest pointwise confidence interval for the principal components is the moment-based, or Wald confidence interval. For the $i^{th}$ element of the $k^{th}$ PC, the moment based CI is defined as $E(\mathbf{V}_{[i,k]}^{b})\pm\sigma(\mathbf{V}_{[i,k]}^{b})z_{(1-\alpha/2)}$, where $\alpha$ is the desired alpha  level, $z_{(1-\alpha/2)}$ is the $100(1-\alpha/2$)$^{th}$ percentile of the standard normal distribution, and the $E$ and $\sigma$ functions capturing the mean and standard deviation of $\mathbf{V}_{[i,k]}^{b}$ across bootstrap samples. Both $E(\mathbf{V}_{[i,k]}^{b})$ and $\sigma(\mathbf{V}_{[i,k]}^{b})$ can be attained without calculating or storing the full bootstrap distribution of $\mathbf{V}_{[i,k]}^{b}$ (see section \ref{sec:Bootstrap-Moments-of}).

Another common pointwise interval for $\mathbf{V}_{[i,k]}^{b}$ is the bootstrap percentile CI, defined as ($q(\mathbf{V}_{[i,k]}^{b},\alpha/2)$, $q(\mathbf{V}_{[i,k]}^{b},1-\alpha/2)$), where $q(\mathbf{V}_{[i,k]}^{b},\alpha)$ denotes the $100\alpha^{th}$ percentile of the bootstrap distribution of $\mathbf{V}_{[i,k]}^{b}$. Unlike the moment based CI, the percentile CI does require calculation of the full bootstrap distribution of $\mathbf{V}_{[i,k]}^{b}.$ 

Estimating the percentile interval tends to require more bootstrap samples (e.g. \textbf{$B$}=1000-2000) than estimating the moment-based interval (e.g. $B$=50-200), as the quantile function is more affect affected by the tails of the bootstrap distribution than the moments are \citep{efron1993introduction}. Interpretation of both these pointwise CIs is discussed further in section \ref{sec:apply-FEB-PCA}

Our methods can also be used to quickly calculate bias corrected and accelerated ($BC_{a}$) CIs \citep{efron1987better}, as others have suggested \citep{timmerman2007estimating}.

\subsubsection{Confidence regions for the principal components\label{sub:CRs-for-PCs}}

Each principal component can be represented as a point in $p$-dimensional space. More specifically, because of the norm 1 requirement for the PCs, the parameter space for the principal components is restricted to the $p$-dimensional unit hypersphere, $S_{p}=\{\mathbf{x}\in R^{p}:\medspace\mathbf{x}'\mathbf{x}=1\}$. To create $p$-dimensional CRs for each PC, \citet{beran1985covariance} suggest so-called confidence cones on the unit hypersphere, of the form
\[
\{\mathbf{x}\in S_{p}:\;|\mathbf{x}'\mathbf{V}_{[,k]}|\geq q(|\mathbf{V}_{[,k]}^{b'}\mathbf{V}_{[,k]}|,\alpha)=q(|\mathbf{A}_{[k,k]}^{b}|,\alpha)\}
\]

Here, $q(a^{b},\alpha)$ is the quantile function denoting the $100\alpha^{th}$ bootstrap percentile of the statistic $a^{b}$. As noted in section \ref{sub:Switching-the-sign}, the calculation of $\mathbf{V}_{[,k]}^{b'}\mathbf{V}_{[,k]}$ can be simplified to $\mathbf{V}_{[,k]}^{b'}\mathbf{V}_{[,k]}=\mathbf{A}_{[k,k]}^{b}$. Geometrically, the dot product condition of this CR is equivalent to a condition on the angle between $\mathbf{x}$ and $\mathbf{V}_{[,k]}$. Note that this CR automatically incorporates the sign adjustments described in section \ref{sub:Switching-the-sign}. \citet{beran1985covariance} provide a theoretical proof for the coverage of CRs constructed in this way. 

It is also possible to create joint confidence bands (jCBs) for the PCs according the method outlined by \citet{crainiceanu2012bootstrap}. However, such bands will also contain vectors that do not have norm 1, and may even exceed 1 in absolute value for a specific dimension. As a result, many vectors contained within the jCBs will not be valid principal components, which complicates interpretation of the jCBs.

\subsubsection{Confidence regions for the principal subspace\label{sub:CRs-for-PS}}

To characterize the variability of the subspace spanned by the first $K$ PCs, also known as the principal subspace, it is not sufficient to simply combine the individual CRs for each PC. This is because the sampling variability of the individual fitted PCs is influenced by random rotations of the fitted PC matrix $\mathbf{V}_{[,1:K]}^{b}$, while the sampling variability of the subspace is not. Similarly, most models whose fit depends on the leading PCs are unaffected by random rotations.

To characterize the sampling variability of the principal subspace, we first note that any bootstrap principal subspace can be defined by the $p\times K$ matrix with columns equal to the leading $K$ PCs. Any such matrix must be contained within the set of all of $p\times K$ orthonormal matrices. This set can be written as the Steifel manifold $M_{K}(R^{p}):=\{\mathbf{X}\in F^{p\times K}:\mathbf{X}'\mathbf{X}=\mathbf{I}_{K}\}$, where $F^{p\times K}$ is the set of all $p\times K$ matrices. To create CRs for the principal subspace, we can use the following generalization of CRs for the individual PCs 
\[
\{\mathbf{X}\in M_{K}(R^{p}):\;||\mathbf{X}'\mathbf{V}_{[,1:K]}||\geq q(||\mathbf{V}_{[,1:K]}^{b'}\mathbf{V}_{[,1:K]}||,\alpha)=q(||\mathbf{A}_{[1:K,1:K]}^{b}||,\alpha)\}
\]

Here, the norm operation refers to the Frobenius norm. \citet{beran1985covariance} suggest CRs of this form to characterize variability of a set of sample covariance matrix eigenvectors whose corresponding population eigenvalues are all equal. However, the CR construction method can also be applied in the context of estimating the principal subspace. As with CRs for the individual PCs, we can make the simplification that $\mathbf{V}_{[,1:K]}^{b'}\mathbf{V}_{[,1:K]}=\mathbf{A}_{[1:K,1:K]}^{b}$. Note that such CRs automatically adjust for random rotations of the first $K$ principal components -- if $\mathbf{R}$ is a $K\times K$ orthonormal transformation matrix, then $||(\mathbf{XR})'\mathbf{V}||=||\mathbf{X}'\mathbf{V}||$.

\subsection{Maintaining informative rotational variability \label{sub:rotational-var}}

When several of the leading eigenvalues of the population covariance matrix are close, the fitted PCs in any sample may be a mixtures the leading population PCs. In these cases, the bootstrap PCs will often be approximate rotations of the leading sample PCs. Others have argued if the parameter of interest is the principal subspace or the model fit, then the bootstrap PCs should be adjusted to correct for rotational variability, as the principal subspace is unaffected by rotations among the leading PCs. Specifically, it has been suggested to use a Procrustean rotation to match the bootstrap PCs to the original sample PCs \citep{milan1995application}, and to then create pointwise confidence intervals (CIs) based on the rotated PCs \citep{timmerman2007estimating,babamoradi2012bootstrap}.%
\footnote{One interpretation of CIs constructed from rotation adjusted bootstrap PCs is that if the population PC matrix is rotated towards the each sample from the population, then average pointwise coverage of rotation adjusted CIs should be approximately $100\alpha\%$%
} We argue however that bootstrap rotational variability is informative of genuine sampling rotational variability, and that adjusting for rotations is not an appropriate way to represent sampling variability of the principal subspace, or the sampling variability of model fit. This is because pointwise CIs are not designed to estimate the sampling variability of the principal subspace. The pointwise CIs generated from rotated bootstrap PCs also do not capture the sampling variability of standard PCs, as the rotated PCs are not valid solutions to the PCA algorithm. 

Rather than rotating towards the sample, it has also been proposed to rotate both the sample and bootstrap PCs towards a $p\times K$ target matrix $\mathbf{T}$, which is pre-specified before collecting the initial sample $\mathbf{Y}$ \citep{raykov1999note,timmerman2007estimating}.%
\footnote{The computational complexity of finding the appropriate rotation matrix in each bootstrap depends on the taking the SVD of the $K\times K$ matrix $\mathbf{V}_{[,1:K]}^{b'}\mathbf{T}=\mathbf{A}_{[,1:K]}^{b'}\mathbf{V}'\mathbf{T}$, where $\mathbf{V}'\mathbf{T}$ can be pre-calculated before the bootstrap procedure.%
} The target matrix $\mathbf{T}$ may be based on scientific knowledge, or previous research. Such an approach can also be used to test null hypotheses about the principal subspace by rotating $\mathbf{V}_{[,1:K]}^{b}$ toward a null PC matrix $\mathbf{V}_{0}$ \citep{raykov1999note}, and checking if elements of $\mathbf{V}_{0}$ are contained in the resulting CRs 

Our opinion is that if investigators are interested in the sampling variability of the output from a model that uses PCA, then it is the model output, and not the principal components, for which CRs should be calculated. If the sampling variability of the subspace is of interest, than CRs specifically designed for the subspace should be used (see section \ref{sub:CRs-for-PS}), rather than adjusted CIs for the elements of the PCs. Rotating towards a pre-specified target matrix $\mathbf{T}$ can also be a useful approach, although it may be more interpretable to calculate the bootstrap distribution of the variance explained by the columns of $\mathbf{T}$,%
\footnote{In each bootstrap sample, the variance explained by the columns of $\mathbf{T}$ is equal to the variance of the resampled data after a projection onto the space spanned by $\mathbf{T}$. The projected data is equal to $\mathbf{T}(\mathbf{T}'\mathbf{T})^{-1}\mathbf{T}'\mathbf{Y}^{b}=(\mathbf{T}(\mathbf{T}'\mathbf{T})^{-1}\mathbf{T}'\mathbf{V})\mathbf{DU}'\mathbf{P}^{b}$, where $\mathbf{T}'(\mathbf{T}'\mathbf{T})^{-1}\mathbf{T}'\mathbf{V}$ is an $n\times n$ matrix that can be precalculated before the bootstrap procedure.%
} rather than the bootstrap distribution of the fitted PCs after a rotation towards $\mathbf{T}$.

\section{Coverage rate simulations \label{sec:Simulations-of-CI}}

In this section we present simulated coverage rates for the CRs described in section \ref{sec:Construction-of-CIs}. In order to make these simulations as realistic as possible, we simulated data using the empirical PC vectors of the EEG dataset as the true population basis vectors. As a baseline simulation scenario we set the sample size ($n$) equal to $392$, and the true number of basis vectors in the population (denoted by $K_{0}$) equal to 5. 

Measurement vectors for each subject were simulated according to the model $\mathbf{y}_{i}=\sum_{k=1}^{K_{0}}s_{ik}\mathbf{\Psi}_{k}+\bm{\epsilon}_{i}$, where $\mathbf{y}_{i}$ is a $p$-length vector of simulated measurements for the $i^{th}$ subject; $\mathbf{\Psi}_{k}$ is the $k^{th}$ true underlying basis vector, which is set equal to the $k^{th}$ empirical PC of the EEG dataset; $s_{ik}$ is a random draw from the empirical, univariate distribution of the scores for the $k^{th}$ PC; and $\bm{\epsilon}_{i}$ is a vector of independent random normal noise variables, each with mean 0 and variance $\sigma^{2}/p$. Setting the variance of ${\bm{\epsilon}_{i}\ensuremath{}}$ equal to $\sigma^{2}/p$ implies that the total variance attributable to the random noise will be approximately equal to $\sigma^{2}$, and will not depend on the number of measurements ($p$). The parameter $\sigma^{2}$ was set equal to the sum of the variances of the $K_{0}+1$ to $n^{th}$ empirical score variables, implying that for each simulated sample, the first $K_{0}$ basis vectors ($\mathbf{\Psi}_{1},\mathbf{\Psi}_{2},...\mathbf{\Psi}_{K_{0}}$) were expected to explain approximately the same proportion of the variance that they explained in the empirical sample. For each simulated subject, $\mathbf{y}_{i}$, the $K_{0}$ score variables $s_{i1},...s_{iK_{0}}$ were all drawn independently. Coverage was compared across 1000 simulated samples. For each simulated sample, the number of bootstrap samples created for estimation ($B$) was set to 1000.

As comparison simulation scenarios, we increased the number of measurements ($p$) to 5000 and to 20000, by interpolating the empirical EEG data and recalculating the principal components and scores. We also compared against simulated sample sizes $(n)$ of 100 and 250. Because much of the variability in fitting principal components is determined by the spacing of eigenvalues in the population, we simulated separate scenarios where the empirical score distribution was scaled so that each basis vector explained half as much variance as the preceding basis vector. In other words, we scaled true population distribution of scores such that the vector of variances of the 5 score variables was proportional to the vector ($2^{4},2^{3},2^{2},2^{1},1)$. The total variance of the first 5 score variables was kept constant across all simulations. We refer to the modified eigenvalue spacing as the ``parametric spacing'' simulation scenario, and refer to the original eigenvalue spacing as the ``empirical spacing'' simulation scenario. Finally, we also simulated scenarios where the total variance due to the random noise ($\sigma^{2}$) was scaled up 50\%, and where it was scaled down by 50\%. Considering all combinations of eigenvalue spacing, random noise level, sample size, and number of measurements, we conducted $2\times3\times3\times3=54$ sets of simulations.

The total elapsed computation time for these 54 simulations was 28 hours. The simulations were run as a series of parallel jobs on an x86-based linux cluster, using a Sun Grid Engine for management of the job queue. As many as 200 jobs were allowed to run simultaneously. Each job required between approximately .5Gb and 2Gb maximum virtual memory, depending on the scenario being simulated.

\subsection{Simulation results}

\begin{figure}[!t]
\includegraphics[scale=0.92]{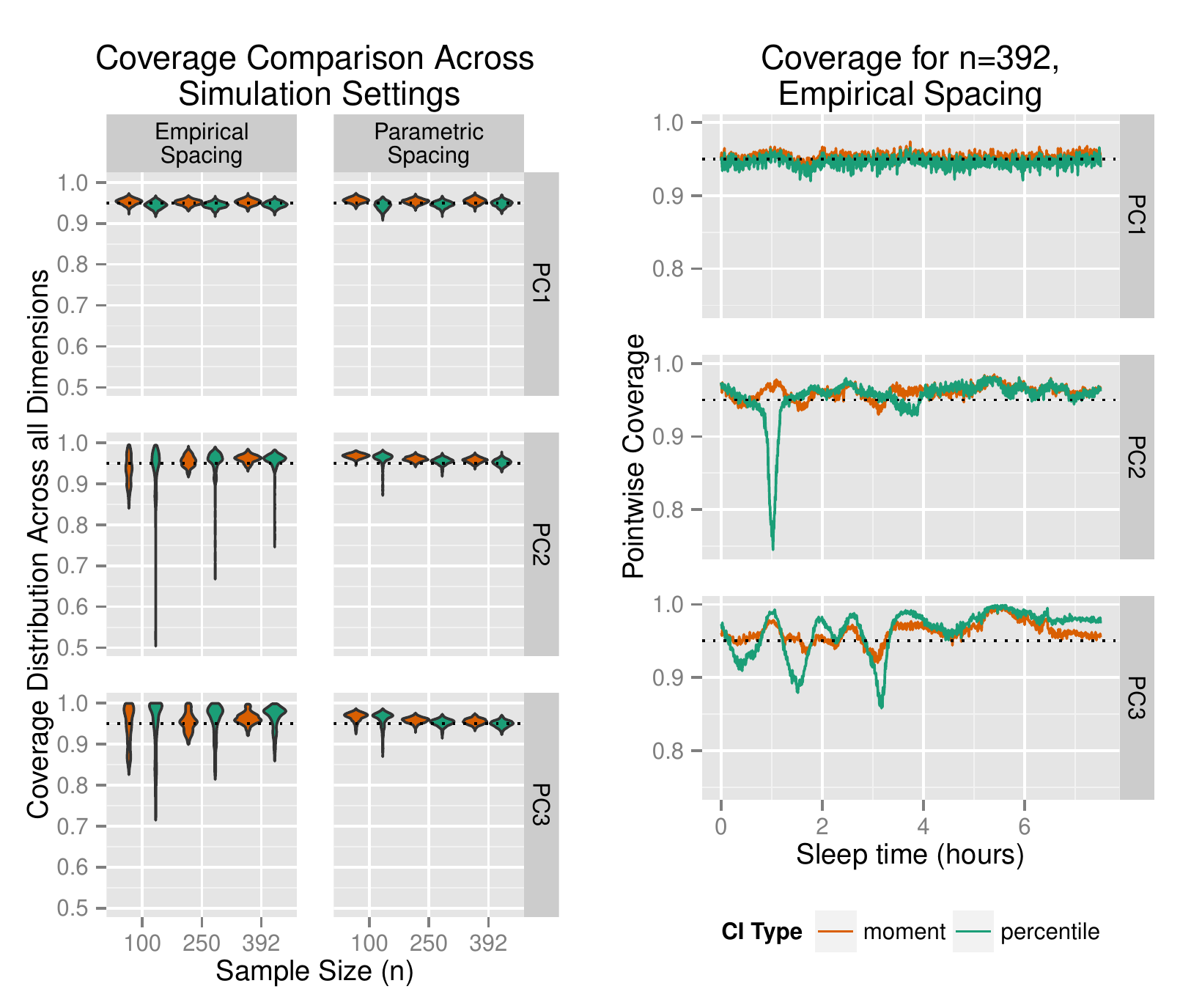}\caption{Pointwise coverage of the PCs - Pointwise bootstrap-based CIs can be calculated for each of the $p$ dimensions of each PC. The violin plots on the left show the distribution of coverage rates across each of the $p$ CIs, under different simulation settings ($p$ fixed at $900$). Simulation cases using the empirical eigenvalue spacing are shown on the left column of violin plots, and simulation cases where where each PC explains half as much variance as the previous PC are shown on the right column. The line plots on the right further explore coverage rates for the specific simulation setting of $n=392$, $p=900$, and the empirical eigenvalue spacing. Coverage rates are shown for each of the $p$ CIs, with the x-axis corresponding to the $p$-dimensional PC element index (time). In both sets of plots, rows correspond to the PC being estimated. \label{fig:coverage-violin+pointwise}}
\end{figure}

We begin by discussing the simulation results for pointwise confidence interval coverage rates in the baseline scenario, with $p=900$, $n=392$, the empirical residual variance, and the empirical eigenvalue spacing. The line plots on the right of Figure \ref{fig:coverage-violin+pointwise} show coverage rates for the each of the $p$ elements of the three PCs. Pointwise coverage for all elements of the first PC is very close to 95\%. For both the second and third PCs, the moment based intervals consistently give close to 95\% coverage, but the percentile intervals appear to give poor coverage in certain regions. This may be an artifact, however, due to how the percentile interval responds to skewness in the underlying bootstrap distribution. Adjusted percentile intervals, such as the $BC_{a}$ interval \citep{efron1987better}, might account for this apparent coverage problem. It is possible that the difficulty in estimating coverage is also affected by the spacing of the eigenvalues -- the first PC corresponds to an eigenvalue that is clearly differentiated, while the eigenvalues for the second two components less clearly differentiated from the remaining eigenvalues.

The violin plots on the left side of Figure \ref{fig:coverage-violin+pointwise} show the distribution of coverage rates across the PC curves as we vary the sample size and eigenvalue spacing. In this panel, the dimensionality ($p$) is fixed at 900, and only the empirical residual noise variance level ($\sigma^{2}$) is used, but results were very similar for alternate levels of dimensionality and residual variance. Coverage rates for all regions of the PC curves converges to 95\% as sample size increases. The coverage is also more accurate when the eigenvalues are well spaced, such as when the first PC is being estimated, or when the parametric spacing for the eigenvalues is used.

\begin{figure}[!t]
\includegraphics[scale=0.89]{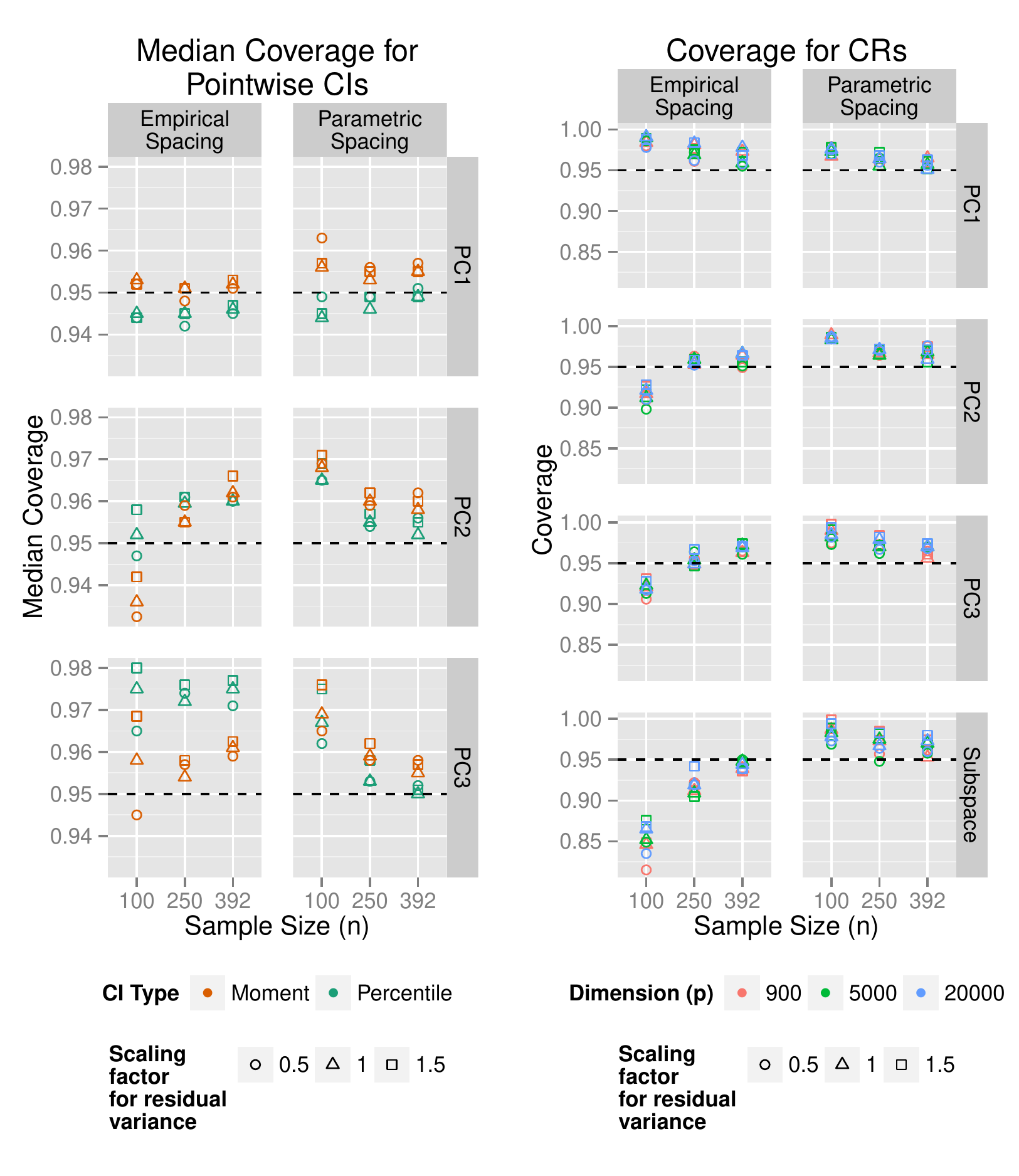}\caption{\label{fig:Cover-median+CR}Coverage across simulation scenarios - The ($3\times2$) array of plots on the left shows the median coverage rate across all $p$ estimated CIs for the PC elements ($p=900$). Rows correspond to the PC being estimated. Simulation cases using the empirical eigenvalue spacing are shown on the left column, and simulation cases where where each PC explains half as much as the previous PC are shown on the right column. The $(4\times2)$ array of plots on the right shows coverage for CRs for the PCs (rows 1, 2 and 3) and for the principal subspace (row 4).}
\end{figure}

The left side of Figure \ref{fig:Cover-median+CR} compares simulation results also across different levels of residual variance. For each combination of sample size ($n$), residual noise ($\sigma^{2}$), and eigenvalue spacing, the median pointwise coverage across all $p$ dimensions is shown for both the moment-based and percentile intervals. In this panel we again fix $p$ at 900, but results were similar for alternate values of $p$. Both the moment-based and percentile intervals generally perform well, with all 54 simulation scenarios having median coverage rates between 93.2\% and 98.1\%. When the eigenvalues of the estimated PCs are well spaced, the coverage rates converge to 95\% as the sample size increases. However, when the eigenvalues are not clearly differentiated, higher sample sizes can lead to slightly overly conservative CIs. 

The right side of Figure \ref{fig:Cover-median+CR} shows the coverage rates of confidence cones for the principal components (section \ref{sub:CRs-for-PCs}) and CRs for the principal subspace (section \ref{sub:CRs-for-PS}). Coverage appears to improve when the eigenvalues are well spaced, and when sample size increases. There does not appear to be a consistent affect of either dimension ($p$) or the scaling factor used for the residual variance.

\section{Applying fast bootstrap PCA\label{sec:apply-FEB-PCA}}

\subsection{Sleep EEG }

When applying fast bootstrap PCA to the EEG dataset, we find that bootstrap estimates of first PC exhibit minimal variability. The second two PCs are estimated with considerably more variability, but most of this variability is due to random rotations among PCs 2 through 4, all of which roughly correspond with oscillatory patterns.

\begin{figure}
\includegraphics[scale=0.5]{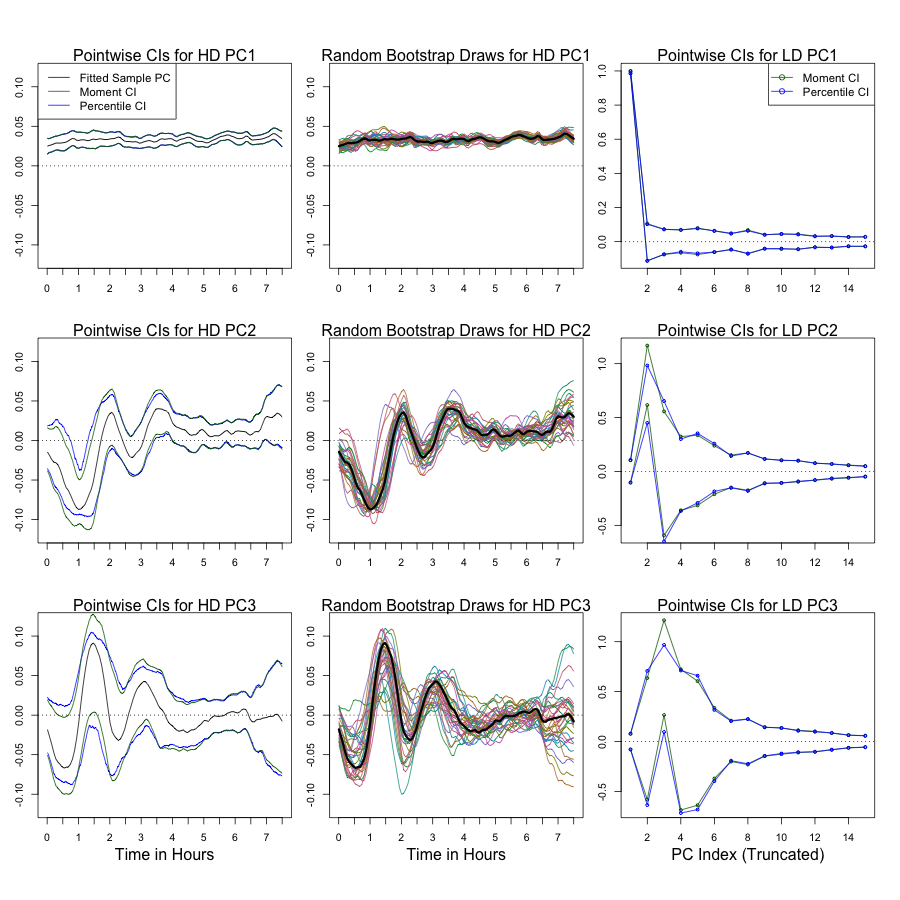}

\caption{Bootstrap variability of principal components - Each row of plots corresponds to a different principal component (PC), either the first, second or third. The left column shows the fitted principal components on the original high dimensional space, along with pointwise confidence intervals. The central column shows the a random selection of 30 draws from the bootstrap distribution each PC. The third column shows pointwise confidence intervals on the low dimensional space, based on the distribution of $\mathbf{A}^{b}$. \label{fig:Boot-PCA-EEG-results}}
\end{figure}

Figure \ref{fig:Boot-PCA-EEG-results} shows the results of this analysis. The left column shows 95\% pointwise intervals for each dimension of each of the three PCs. We see that the moment-based and percentile intervals generally agree, although they tend to differ more when the fitted PC elements are further from zero. Since the width of the percentile and moment-based CIs are fairly similar, disagreements between the two types of intervals are reflective of skewness in the underlying bootstrap distribution. 

The sets of pointwise intervals shown in the left column of Figure \ref{fig:Boot-PCA-EEG-results} form bands around the fitted sample PCs. It's important to note these bands are only calibrated for pointwise 95\% coverage -- they are not expected to simultaneously contain the true population PC in 95\% of samples. Statements about the overall shape of the population PCs that are based on these intervals will be somewhat ad hoc. Furthermore, many curves contained within these bands do not satisfy the norm 1 requirement for principal components, and are not valid solutions to PCA. For example, the upper and lower boundaries of the bands do not have norm 1, and thus are not in the parameter space for the PCs. Similarly, the zero vector is also not in the parameter space. 

Figure \ref{fig:Boot-PCA-EEG-results} shows that both sets of intervals around the first PC vector are fairly tight, implying that there is little sampling variability in the first PC. The pointwise CIs for the 2nd PC are wider, especially in the first four hours of the night, which might erroneously lead readers to think that the oscillatory pattern in $\mathbf{V}_{[,2]}$ is artificial. A similar pattern in bootstrap variability observed for the third PC.

However, a closer inspection of the bootstrap variability gives more evidence of an oscillatory pattern for the second PC. The central column of Figure \ref{fig:Boot-PCA-EEG-results} shows a random set of 30 bootstrap draws for each PC (i.e. the first panel shows $\mathbf{V}_{[,1]}^{b}$ for \textbf{$b=1,...30$}), with the fitted sample PCs overlaid in black. For the first PC we again see very little bootstrap variation. For the second PC though, we see that the negative spike in hour 1, and the positive spike in hour 2, are often shifted in bootstrap samples. While the pointwise intervals in the left column of Figure \ref{fig:Boot-PCA-EEG-results} seem to indicate the magnitude of these spikes might be lower in the population PCs, the plot of bootstrap draws gives us more information. It shows that the pointwise variability in the oscillatory pattern is more aptly explained by a simultaneous shift of both peaks than by a magnitude change in either peak. The bootstrap draws of $\mathbf{V}_{b[,2]}$ which are most shifted tend to bear a closer resemblance to the third principal component. Bootstrap draws for the third PC do not show as clear a pattern in variation.

The pattern in variation for the second PC can be more succinctly described by noting that the majority of its bootstrap variation is due to random rotations of the second third PCs. Rotational bootstrap variance is shown more explicitly in the third column of Figure \ref{fig:Boot-PCA-EEG-results}, which displays pointwise CIs for the low dimensional bootstrap PC vectors $\mathbf{A}_{[,k]}^{b}$, for $k=1,2,$ and 3. In the first panel we see that the first low dimensional PC vector ($\mathbf{A}_{[,1]}^{b}$) tends to weight highly on the first PC coordinate vector $\mathbf{V}_{[,1]}$, and minimally on the other PCs. The low dimensional second PC ($\mathbf{A}_{[,2]}^{b}$) generally places high weight on $\mathbf{V}_{[,2]}$, but can also place high weight on $\mathbf{V}_{[,3]}$. The CI plot for $\mathbf{A}_{[,3]}^{b}$ reveals that most of the variation in $\mathbf{V}_{b[,3]}$ is in the direction of the either $\mathbf{V}_{[,2]}$, $\mathbf{V}_{[,4]}$, or $\mathbf{V}_{[,5]}$. This pattern is not clear from either of the first two columns of Figure \ref{fig:Boot-PCA-EEG-results}. This rotational variability is a highly relevant aspect of the sampling variability of PCs. 

Note that the moment-based CIs shown on the right column of Figure \ref{fig:Boot-PCA-EEG-results} can exceed one in absolute value, which will surely violate the norm condition for PCs. In practice, such violations should be accounted for by truncating the CIs at -1 and 1, but we keep the violation for illustrative purposes in Figure \ref{fig:Boot-PCA-EEG-results}. It is also worth noting that the percentile CIs for $\mathbf{A}_{[k,k]}^{b}$ will rarely include the value 1, which can be thought of as the fitted value of $\mathbf{A}_{[k,k]}^{b}$ in the original sample. The low dimensional percentile CIs for the elements of $\mathbf{A}^{b}$ also fully contain the information required to create confidence cones for each PC (section \ref{sub:CRs-for-PCs}).

\begin{figure}
\includegraphics[scale=0.85]{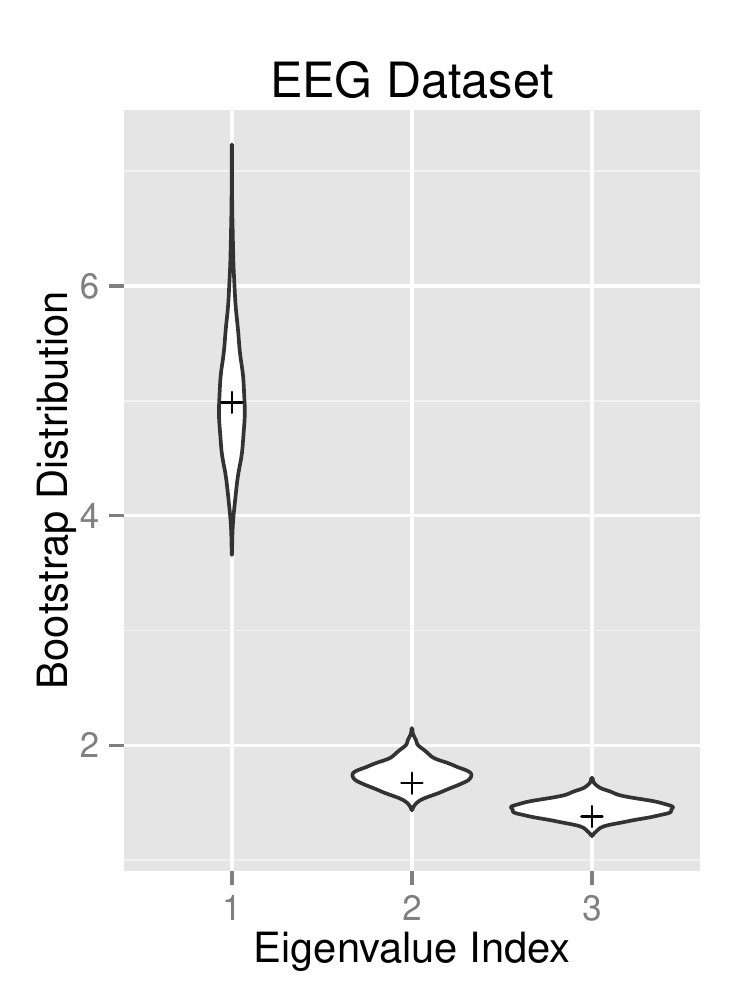}\includegraphics[scale=0.85]{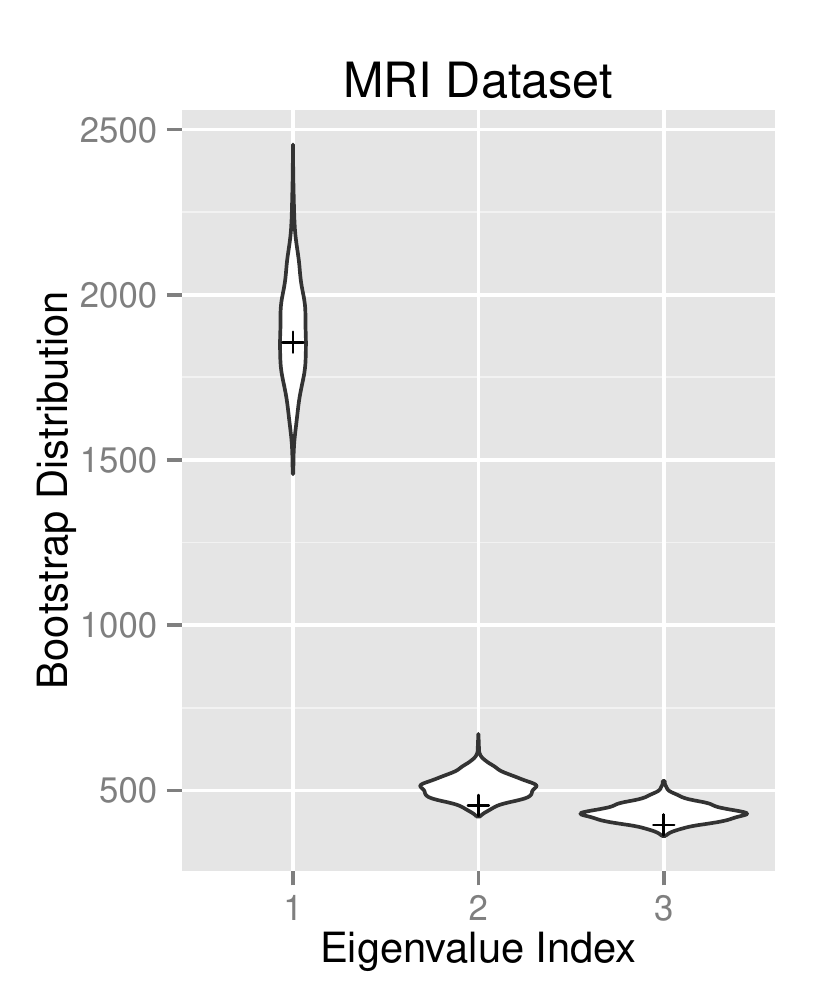}\caption{Bootstrap eigenvalue distribution - For both the EEG and MRI datasets, we show bootstrap distribution for the first three eigenvalues of the sample covariance matrix. Tick marks show the eigenvalues from the original sample covariance matrix. \label{fig:Boot-eigenvalue-dist-EEG-MRI}}
\end{figure}

Figure \ref{fig:Boot-eigenvalue-dist-EEG-MRI} shows the bootstrap distribution of the first three eigenvalues of the sample covariance matrix (the diagonals of $(1/(n-1))(\mathbf{D}^{b})^{2}$). In general, there is a known upward bias in the first eigenvalue of the sample covariance matrix, relative to the first eigenvalue of the population covariance matrix \citep{daudin1988stability}. The amount of bias can be estimated using bias in the bootstrap distribution of covariance matrix eigenvalues. Each bootstrap sample can be seen as a simulated draw from the original sample, in which the eigenvalues are known. Here, we define the percent bias in the bootstrap eigenvalues as the difference between the average eigenvalue across all bootstrap samples and the eigenvalue in the original sample, divided by the eigenvalue of the original sample. For the first three covariance matrix eigenvalues in the EEG dataset (Figure \ref{fig:Boot-eigenvalue-dist-EEG-MRI}), there is only a slight upward bias in the bootstrap estimates (percent bias = 1.1\%, 4.5\%, and 5.0\% respectively).

\subsection{Brain MRIs\label{sub:FEB-Brain-Magnetic-Resonance}}

We also apply our bootstrap procedure to estimate sampling variability of the PCs from the brain MRI dataset. This is primarily included as an example to show the computational feasibility of our method in the high dimensional setting. A deeper interpretation of the sample PCs is provided by \citep{zipunnikov2011multilevel_RAVENS,zipunnikov2011ravenspca}.

\begin{figure}[!t]
\begin{centering}
\includegraphics[scale=0.37]{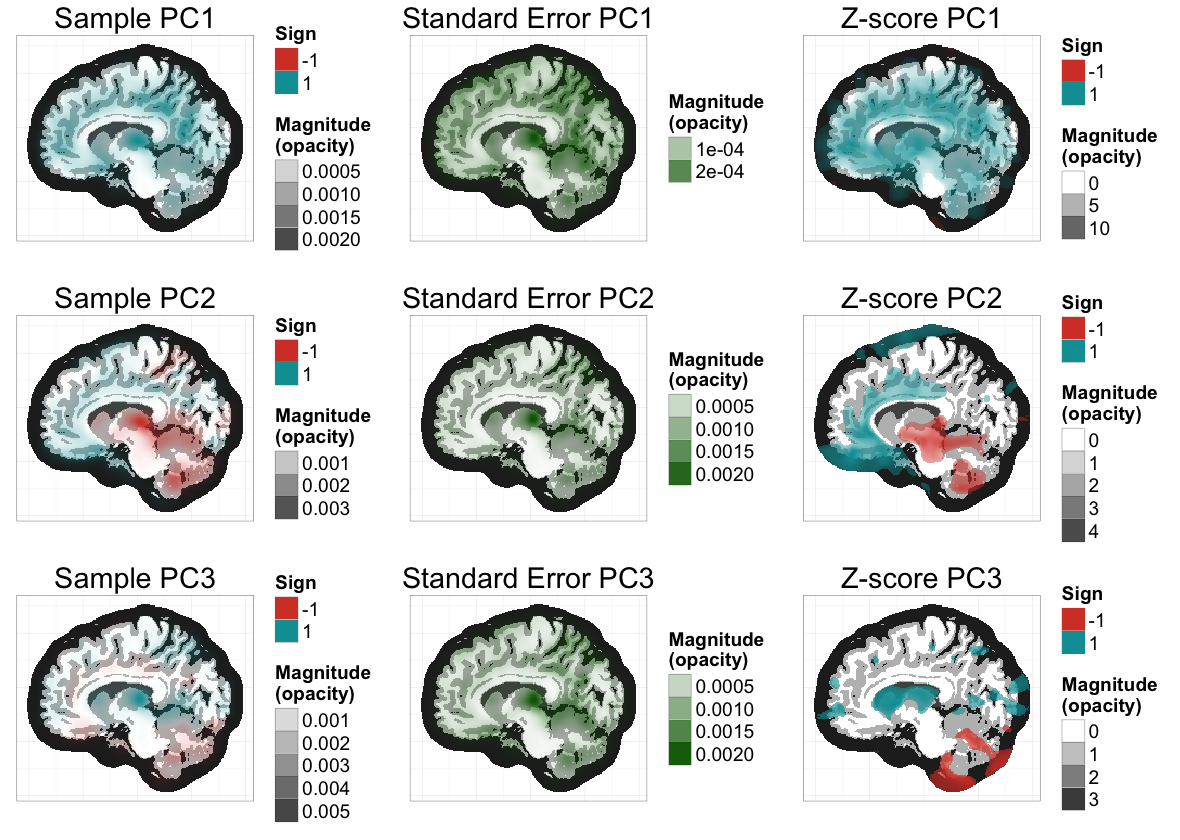}
\par\end{centering}

\caption{Fitted sample values, bootstrap standard errors, and Z-scores for the MRI PCs \label{fig:MRI-PC-se-zscore}}

\end{figure}

Our results imply that the first two sample PCs are estimated with fairly low sampling variability, but that sampling variability is higher for the third PC. The first two columns of figure \ref{fig:MRI-PC-se-zscore} respectively show the fitted sample PCs and the bootstrap standard errors for the PCs. The standard errors are generally of a lower order of magnitude than the corresponding fitted values for the PCs. A direct comparison is given in the third column of Figure \ref{fig:MRI-PC-se-zscore}, which shows the fitted sample PCs divided by their pointwise bootstrap standard errors. These ratios can be interpreted as Z-scores under the element-wise null hypotheses that the population value of any one element of the population PC is zero. Z-scores with absolute value less than 1.96 are omitted from the display. 

\begin{figure}
\includegraphics[scale=0.35]{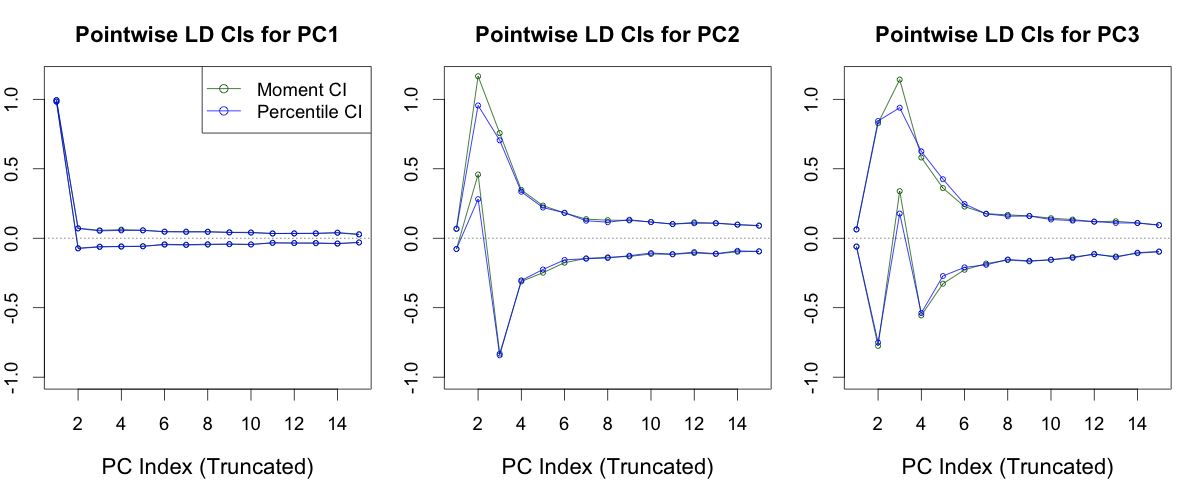}

\caption{\label{fig:MRI-PC-rotational-Ab-CI}Low dimensional CIs for the MRI PCs - Moment-based and percentile confidence intervals for $\mathbf{A}_{[1:15,k]}^{b}$, where $k=1,2$ and 3.}

\end{figure}

To estimate sampling variability due to rotations of the leading population PCs, Figure \ref{fig:MRI-PC-rotational-Ab-CI} shows pointwise confidence intervals for the truncated vectors $\mathbf{A}_{[,k]}^{b}$, for $k=1,2,3$. These intervals are analogous to the intervals shown on the right of Figure \ref{fig:Boot-PCA-EEG-results}. A substantial proportion of the bootstrap variability for the second two PCs is due to random rotations between them.

The second panel of Figure \ref{fig:Boot-eigenvalue-dist-EEG-MRI} shows the bootstrap distribution of the eigenvalues of sample covariance matrix. Relative to the fitted eigenvalues in the original sample, the bootstrap eigenvalues show a small, but notable upward bias (percent bias = 1.7\%, 12.2\%, and 9.2\% respectively).

\subsection{Computation times for bootstrap PCA\label{sub:Calculation-times-for}}

We tested the speed of our bootstrap PCA procedure for several combinations of sample size ($n$) and dimensionality $(p$). Varying $n$ and $p$ was achieved by using subsets of the measurements and subjects from the MRI dataset (section \ref{sub:Brain-Magnetic-Resonance}). All calculations were run on a standard laptop (2.5GHz Intel Core i5, 12 Gb memory), without parallelization.

\begin{figure}[!t]
\includegraphics[scale=0.6]{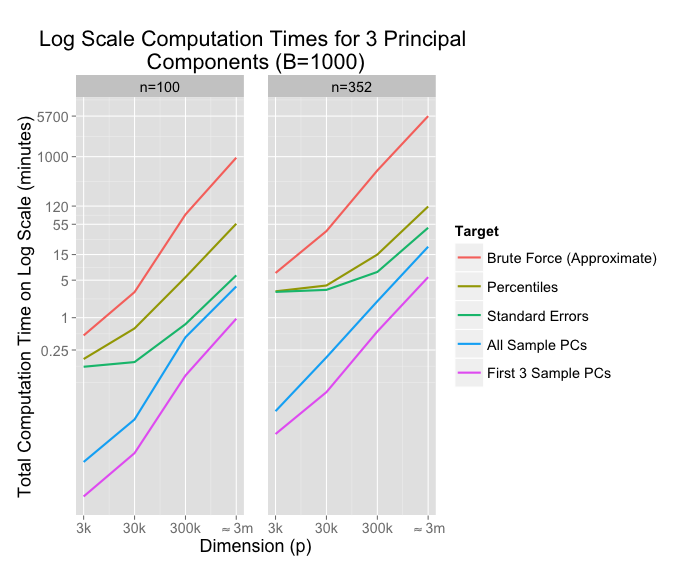}

\caption{Computation times for bootstrap PCA \label{fig:Compution-Times} - The two plots show computation times for sample sizes of 100 (left) and 352 (right). The horizontal axis shows the dimensionality ($p$ = 3,000; 30,000; 300,000; and 2,979,666) and the vertical axis shows total elapsed computation time of each method. The spacing for both axes is on the log scale, in base 10. Computation times are shown for calculating the first 3 sample PCs, all $n$ sample PCs, bootstrap standard errors, and bootstrap percentiles. For the bootstrap standard errors and percentiles, the computation time shown includes the time required for the full SVD of the original sample. An approximation of the time required to calculate the bootstrap distribution of the PCs using standard methods is also shown.}
\end{figure}

Figure \ref{fig:Compution-Times} shows the results of these tests. We compare our proposed methods against an approximate ``brute force'' calculation time, which is attained by multiplying the calculation time for the first 3 sample PCs by the number of bootstrap samples ($B=1000$). This approximation is conservative in that it does not include time required for saving and loading the $p$-dimensional bootstrap PCs. Still, our methods offer significant speed improvements over the approximate brute force method in all tested scenarios. In particular, for the most computationally demanding scenario tested ($p$ = 2,979,666; $n=352$), pointwise percentile intervals based on the full bootstrap distribution of the PCs were calculated in 118 minutes using our method, as opposed to 5,693 minutes (3.95 days) with the brute force method. Calculating bootstrap standard errors with our method (section \ref{sec:Bootstrap-Moments-of}) took only 47 minutes.

While the brute force method can be parallelized on a high powered computing cluster to reduce the total elapsed calculation time, the parallelization procedure will incur bottlenecks when multiple nodes attempt to simultaneously load the sample data files into memory. The sample data files will only be able to be accessed by one node at a time. This is an especially relevant problem for the high dimensional scenario, when the data must be stored as a set of block matrices that are loaded into memory sequentially (section supplemental materials). In contrast, our proposed method for fast, exact bootstrap PCA can be parallelized without incurring these bottlenecks, as each node only needs to import the $n\times n$ matrix of sample scores ($\mathbf{DU}'$).

\section{Discussion\label{sec:Discussion}}

In this paper we outline methods for fast PCA in high dimensional bootstrap samples, based on the fact that all bootstrap samples lie in the same low dimensional subspace. We show computational feasibility by applying this method to a sample of sleep EEG recordings ($p=900$), and to a sample of processed brain MRIs ($p=$2,979,666). Bootstrap standard errors for the first three components of the MRI dataset were calculated on a commercial laptop in 47 minutes.

Ultimately, the usefulness of high dimensional bootstrap PCA will depend not on its speed, but on its demonstrated ability to capture sampling variability. We found that the bootstrap performed well in the simulation settings presented here (section \ref{sec:Simulations-of-CI}). However, bootstrap PCA has rarely been applied to high dimensional data in the past, and its theoretical properties in high dimensions are still not well studied. This lack of study is likely due to the computational bottlenecks of standard bootstrap PCA, which are compounded in theoretical research that includes simulation studies. Our hope is that the methods presented here will expand the use of bootstrap PCA, and allow for theoretical properties of the bootstrap PCA procedure to be studied and verified via simulation.

When interpreting the results of bootstrap PCA, we find it particularly useful to generate confidence intervals around elements of the low dimensional $\mathbf{A}^{b}$ matrices (Figures \ref{fig:Boot-PCA-EEG-results} and \ref{fig:MRI-PC-rotational-Ab-CI}). These CIs are a parsimonious way to display the dominant directions in PC bootstrap variability, which often correspond to rotations among the leading sample PCs. Calculating these CIs also does not require operations on the $p$-dimensional scale, beyond the initial SVD of the sample.

One alternative potential method for describing the dominant patterns in bootstrap PC variability, is to use $p$-dimensional elliptical CRs of the form 
\[
\{\mathbf{x}\in S_{p}:\,\:(\mathbf{x}-\mathbf{V}_{[,k]})'Cov(\mathbf{V}_{[,k]}^{b})^{-}(\mathbf{x}-\mathbf{V}_{[,k]})\leq q((\mathbf{V}_{[,k]}^{b}-\mathbf{V}_{[,k]})'Cov(\mathbf{V}_{[,k]}^{b})^{-}(\mathbf{V}_{[,k]}^{b}-\mathbf{V}_{[,k]}),\alpha)\}
\]
Where $Cov(\mathbf{V}_{[,k]}^{b})$ is the $p\times p$ bootstrap covariance matrix of the $k^{th}$ PC, and $Cov(\mathbf{V}_{[,k]}^{b})^{-}$ is the generalized inverse of $Cov(\mathbf{V}_{[,k]}^{b})$. Note that the use of the generalized inverse, or some form of regularization, is required, as the covariance matrix $Cov(\mathbf{V}_{[,k]}^{b})$ is not full rank, and not invertible. As a result, these regions will not describe sampling variability in directions orthogonal to the span of the observed sample points. Note also that $Cov(\mathbf{V}_{[,k]}^{b})^{-}=(\mathbf{V}Cov(\mathbf{A}_{[,k]}^{b})\mathbf{V}')^{-}=\mathbf{V}(Cov(\mathbf{A}_{[,k]}^{b})^{-})\mathbf{V}'$. Thus, the above CR is equivalent to the easily calculable region

\[
\{\mathbf{x}\in S_{p}:\,\:(\mathbf{V}'\mathbf{x}-\delta_{k})'Cov(\mathbf{A}_{[,k]}^{b})^{-}(\mathbf{V}'\mathbf{x}-\delta_{k})\leq q((\mathbf{A}_{[,k]}^{b}-\delta_{k})'Cov(\mathbf{A}_{[,k]}^{b})^{-}(\mathbf{A}_{[,k]}^{b}-\delta_{k}),\alpha)\}
\]

Where $\delta_{k}$ is the $k^{th}$ column of the $n\times n$ identity matrix. These elliptical CRs can be fully defined by the length and directions of their primary axes, which, in the case of spacial data, can be plotted on the p-dimensional scale. 

Interpretation of bootstrap PCA results is complicated by the fact that many PCA results are interdependent. For example, each PC is only defined conditionally on the preceding PCs. If we want to isolate only the variability of the $k^{th}$ PC that affects this conditional interpretation, it can be useful to first assume that the first $k-1$ PCs are estimated without error. Logistically, we can condition on the leading $k-1$ PCs by resampling from the residuals after projecting the dataset onto the matrix $\mathbf{V}_{[,1:(k-1)]}$. This is equivalent to setting the first $k-1$ score variables to zero before starting the resampling process. Alternatively, we could assume that the first PC is a mean shift, and estimate the sampling variability of the remaining PCs by resampling from the residuals after projecting the dataset onto a constant, flat basis vector. This general approach requires the strong assumption that the leading PCs are known, but the procedure can still be useful in exploring the sources of PC variability.

\section*{Acknowledgements}

The methodological research presented here was supported by the National Institute of Environmental Health Sciences (grant number T32ES012871), the National Institute of Biomedical Imaging And Bioengineering (grants RO1 EB012547 and P41 EB015909), and the National Institute of Neurological Disorders and Stroke (grant RO1 NS060910). Recording and maintenance of the MRI dataset was supported by the National Institute on Aging (grant R01 AG10785). The content of this article is solely the responsibility of the authors.

\section*{R Package Code}

Code for this paper is available as an R package at

\url{https://github.com/aaronjfisher/bootSVD}

{\setstretch{1.0}

\bibliographystyle{apalike}
\bibliography{bootSvd}

}
\end{document}